# Magnetic Properties of Bismuth Ferrite Nanopowder Obtained by Mechanochemical Synthesis

Izabela Szafraniak-Wiza[a*], Bartłomiej Andrzejewski[b], Bożena Hilczer[b]

[a] Institute of Materials Science and Engineering,
Poznań University of Technology,
M. Skłodowska-Curie Sq. 5, PL-60965 Poznań, Poland
* corresponding author: izabela.szafraniak-wiza@put.poznan.pl

[b] Institute of Molecular Physics
Polish Academy of Sciences
Smoluchowskiego 17, PL-60179 Poznań, Poland

*Abstract* — **Multiferroic bismuth ferrite (BiFeO$_3$) nanopowders have been obtained in room temperature by mechanical synthesis. Depending on the post-synthesis processing the nanopowders have exhibited differences in the mean sizes, presence of amorphous layer and/or secondary phases. Extended magnetic study performed for fresh, annealed and hot-pressed nanopowders have revealed substantial improvement of the magnetic properties in the as-prepared powder.**

## I.  INTRODUCTION

Multiferroics exhibit at least two primary ferroic orders: ferroelectric, ferromagnetic, ferroelastic or ferrotoroic in a single homogeneous phase and these order parameters can be mutually coupled [1]. Especially interesting are ferroelectromagnets (or magnetoelectric multiferroics) having magnetization and dielectric polarization, which can be modulated and activated by an external electric field and magnetic field, respectively. For this reason, multiferroic materials are being considered for a host of potential applications like magnetic recording media, information storage, spintronics, and sensors (for review of physics and applications of multiferroics see ref. [2]). Between multiferroics bismuth ferrite (BiFeO$_3$) attacked the most scientific interest because its antiferromagnetic and ferroelectric properties are observed at room temperature and the phase transitions occur at very high temperatures *i.e.* $T_N$~640 K and $T_C$~1100 K [2]. At room temperature BiFeO$_3$ has rhombohedrally distorted cubic perovskite cell (*R3c*) and the antiferromagnetic properties are related to G-type ordering with a cycloid modulation (62 nm) apparent down to 5 K.

Recently it has been shown that BFeO$_3$ nanoparticles exhibit strong size-dependent magnetic properties. That effect has been correlated with: (1) increased suppression of the known spiral spin structure with decreasing nanoparticle size, (2) uncompensated spins with spin pinning and strain anisotropies at the surface [3] and (3) presence of impurities (like γ-Fe$_2$O$_3$) and/or oxygen vacancies [4]. In the current work, we studied magnetic properties of BiFeO$_3$ nanopowders with different sizes obtained by room temperature mechanochemical synthesis [5, 6], high-temperature annealing and bulk ceramics made from this powder by hot-pressing method.

## II.  EXPERIMENTAL

Bismuth ferrite nanopowder was synthesized by mechanochemical route. Details of synthesis were published in previous paper [6]. Commercially available oxides (Bi$_2$O$_3$ and Fe$_2$O$_3$ purchased from Aldrich, 99% purity) in stoichiometric ratio were milled in a SPEX 8000 Mixer Mill for 120 h. The thermal treatment was performed for 1h in normal atmosphere at 500 $^o$C. The ceramic samples were prepared by hot isostatic pressing of nanopowder at 200 MPa and 800 $^o$C for 2h. Magnetic properties were studied by Oxford Instruments Ltd. MagLab 2000 System in temperature range between 2÷350 K. The temperature dependent measurements were carried out in two different procedures: zero field cooled (ZFC) and field cooled (FC) one. In the case of ZFC measurements the sample was first cooled in absence of external magnetic field and next the field was applied when the sample reached desired temperature. For FC measurements the sample was cooled in applied magnetic field (1 mT÷1 T). In both cases data were acquired during the heating cycle.

## III.  RESULTS AND DISCUSSION

Earlier XRD studies [6] confirm that the nanopowder obtained after 120 h milling exhibits rhombohedrally distorted perovskite structure. The mean grain size estimated using Scherrer formula is 19÷26 nm. Transmission electron microscope studies (TEM) indicate that nanograins of this powder are composed of crystalline core BFO and amorphous shell [6]. Thermal annealing (1h at 500 °C) causes crystallization of the shell and his leads to the increase of the mean grain size to 26÷42 nm. This sample contains also substantial amount of $Bi_2Fe_4O_9$ parasitic phase formed during high temperature processing. Hot pressing method yields dense, bulk bismuth ferrite ceramic.

These three different $BiFeO_3$ samples were used for magnetometric measurements. ZFC and FC temperature dependences of magnetic susceptibility $\chi(T)$ were measured for all samples and for various applied magnetic fields. The origin of the difference between ZFC and FC susceptibilities in as-prepared nanopowder is domain wall pinning and/or reorientations of weak ferromagnetic domains in bismuth ferrite phase. A spin-glass phase as a source of this difference should be rather excluded because in BFO compounds there is a lack of memory, which is common effect for spin-glass systems [7]. In ZFC susceptibility anomalous, field-dependent maximum about 8 K (for the field 1 T) is observed whereas FC susceptibility decreases monotonously with temperature (see Fig. 1).

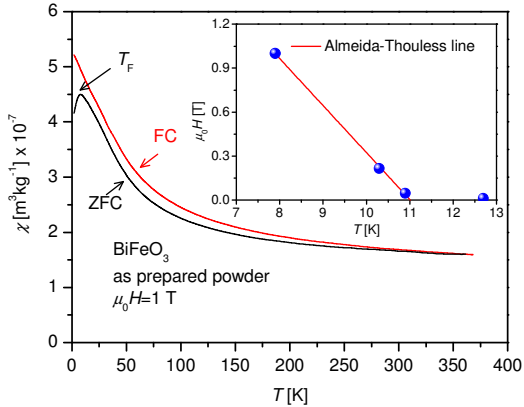

Figure 1. ZFC and FC magnetic susceptibilities for as-prepared nanopowder. The insert shows freezing temperature dependence on magnetic field fitted with Almeida-Thouless model.

The low temperature maximum is probably related to spin clusters in amorphous shell of nanograins [8]. According this approach, at high temperature the spins are in a paramagnetic state. At lower temperature the ordered spin clusters are formed, which freeze into random directions below temperature $T_F$. It is assumed that the freezing temperature $T_F$ corresponds to the maximum of ZFC susceptibility. The validity of the above explanation can be tested using the Almeida-Thouless relation between the freezing temperature and the applied magnetic field [8]:

$$H = H_0 \left[ 1 - \frac{T_F(H)}{T_F(0)} \right]^{3/2}. \quad (1)$$

In eq. (1) $H_0$ denotes magnetic field for which $T_F \to 0$ and $T_F(0)$ is the freezing temperature for $H=0$. The best fit using eq. (1) to the data is obtained for the parameters: $\mu_0 H_0 = 3.5$ T and $T_F(0) = 11$ K (see the inset to Fig. 1).

The annealed sample shows no low temperature anomaly but it exhibits substantial difference between ZFC and FC susceptibilities (Fig. 2a). The absence of maximum in ZFC susceptibility well corresponds to the crystallization of amorphous shell and thus disappearance of a spin-glass phase after thermal annealing. The thermal processing also leads to the increase of the mean grain size to 26÷42 nm. In these nanoparticles with improved crystal structure we expect enhanced energy barriers for magnetic moment reorientations or pinning of domain walls leading to pronounced irreversibility effects. The anomaly observed about 250 K is very common for orthoferrites [2]. However, the calcined sample contains $Bi_2Fe_4O_9$ parasitic phase which undergoes a transition from paramagnetic to antiferromagnetic state near $T_N = 264 \pm 3$ K [9] which can be also produce this anomaly. For the hot-pressed sample, the ZFC and FC susceptibilities (Fig. 2b) are almost identical, with pronounced anomaly about 250 K. This indicates weak pinning of magnetic domain or domain walls in hot-pressed sample.

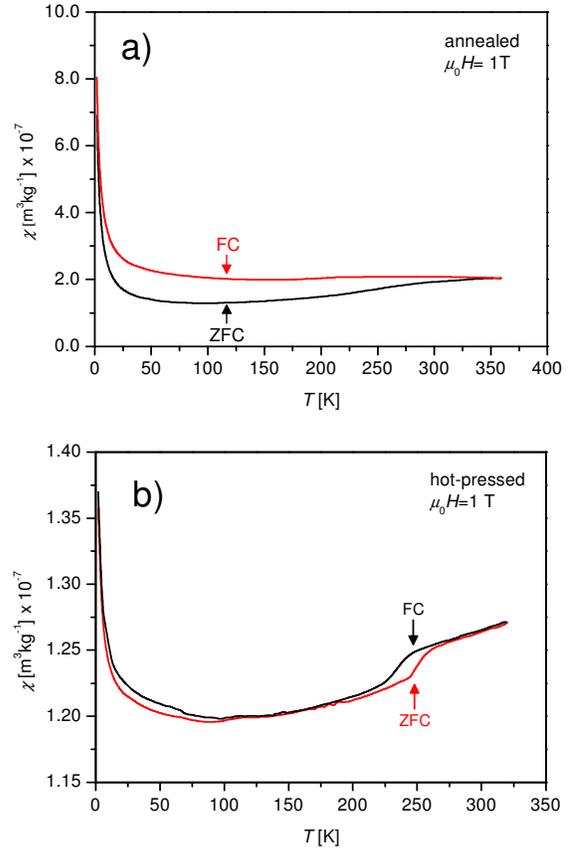

Figure 2. ZFC and FC magnetic susceptibilities for calcined BFO nanopowder (a) and hot-pressed ceramics (b). The applied magnetic field was equal to 1 T.

The magnetic hysteresis loops presented in Fig. 3 have been recorded at temperatures 4 K, 50 K and 300 K for all samples.

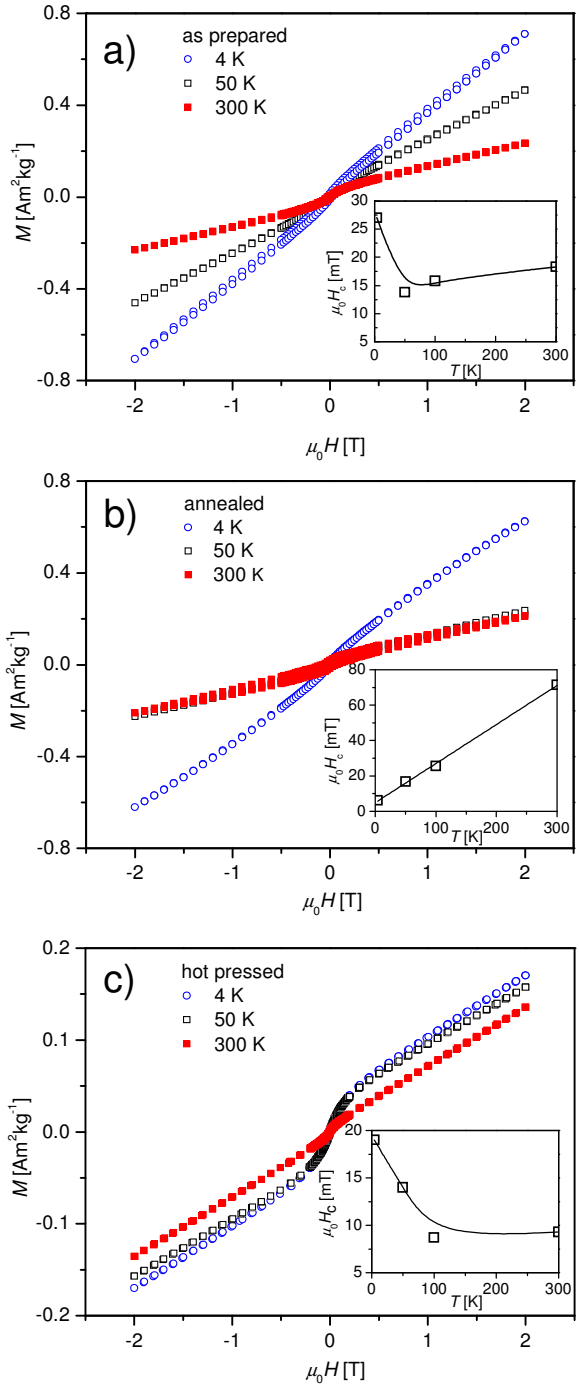

Figure 3. Hysteresis loops recorded at different temperatures for a) as-prepared b) annealed nanopowders and c) hot-pressed ceramics made from nanopowder. Insets show temperature dependence of coercive field for each sample.

The hysteresis loops for nanopowders are not well-saturated which may be related either to antiferromagnetic state or to the spin disorder and spin pinning at the nanograin surfaces [3]. The maximum of magnetization, $M_S$=0.710 Am$^2$/kg is observed at 4 K and 2 T for the as-prepared nanopowders composed of shell-core nanograins (see Fig. 3a). This value is, depending on temperature range, 2÷4 times bigger than $M_S$ for the thermal processed ceramics. The increased magnetization is due to size effect and presence of amorphous shell. All recorded loops are superposition of ferromagnetic non-linear and antiferromagnetic or paramagnetic linear contributions. The nonlinear, ferromagnetic contribution is most pronounced in low temperatures and for as-prepared BFO sample (Fig. 3a). This ferromagnetic behavior can be related to suppression of spin cycloid leading to the lack of spin compensation or to surface spins in BFO nanograins.

The complex relation of coercive field $H_c(T)$ on temperature found in the BFO samples has different origins. For the as-prepared BFO sample (inset to Fig. 3a) the coercivity $H_c(T)$ exhibits non-monotonic behaviour with the highest value over 25 mT at the temperature 4 K and minimum at 50 K. This dependence is a superposition of strong pinning of spin clusters below freezing temperature $T_F$ and pinning of weak ferromagnetic domains prevailing at higher temperatures. The temperature dependence of coercivity for calcined BFO sample (inset to Fig. 3b) is also unusual because $H_c(T)$ linearly increases with temperature. The increase of coercivity with temperature sometimes takes place in granular systems if intergranular coupling decreases with temperature [10]. In this case, the coercivity $H_c(T)$ is described by a phenomenological formula:

$$\mu_0 H_c = \alpha_{ex}\alpha_k \mu_0 H_a - N_{eff} J_s. \qquad (2)$$

where: $\alpha_k$ is the structure-dependent Kronmuller parameter, $a_{ex}$ parameter represents deleterious influence of the intergrain coupling, $N_{eff}$ is the magnetostatic parameter, $J$ is the magnetization and $H_a$ denotes the anisotropy field. For the nanopowders the intergranular coupling is obviously very week. This coupling is temperature dependent and can be further suppressed at higher temperatures. For dense, hot-pressed ceramic the intergranular coupling is very strong and the coercivity reflects only decrease of anisotropy field $H_a(T)$ with temperature (see inset to Fig. 3c).

## IV. CONCLUSIONS

The studies of the magnetic properties of BiFeO$_3$ nanopowder obtained by mechanochemical synthesis and its ceramics have shown the improvement of magnetization. The improvement in the magnetization of nanosized particles we would like to relate to the suppression of cycloidal order, *i.e.*, incomplete rotation of the spins along the direction of the wave vector and also to an increase in spin canting, due to the lattice strain which gives rise to weak ferromagnetism. The presence of the core-shell structure in the BiFeO$_3$ grains obtained by mechanochemical synthesis may be important from the application point of view because of possibility to improve magnetic properties of nanomaterials.

V. ACKNOWLEDGEMENT

Authors would like to thank Dr. B. Malic (Institut Jozef Stefan, Ljubljana, Slovenia) for preparation of the hot-pressed ceramics.